# Bragg soliton compression and fission on a CMOS-compatible platform

Ezgi Sahin[1,2,*], Andrea Blanco-Redondo[3,4], Peng Xing[1], Doris K. T. Ng[5], Ching E. Png[2], Dawn T. H. Tan[1] and Benjamin J. Eggleton[3,4]

[1] Photonics Devices and Systems Group, Singapore University of Technology and Design, 8 Somapah Rd., Singapore 487372, Singapore.
[2] Electronics and Photonics Department, Institute of High Performance Computing, Agency for Science, Technology and Research, 1 Fusionopolis Way, Singapore 138632, Singapore.
[3] Institute of Photonics and Optical Science (IPOS), School of Physics, The University of Sydney, New South Wales 2006, Australia.
[4] The University of Sydney Nano Institute (Sydney Nano), The University of Sydney, New South Wales 2006, Australia.
[5] Institute of Microelectronics, A*STAR, 2 Fusionopolis Way, #08-02, Innovis Tower, Singapore 138634, Singapore.
*ezgi_sahin@mymail.sutd.edu.sg

**Higher-order soliton dynamics, specifically soliton compression and fission, underpin crucial applications in ultrafast optics, sensing, communications and signal processing. Bragg solitons exploit the strong dispersive properties of periodic media near the photonic band edge, enabling soliton dynamics to occur on chip-scale propagation distances and opening avenues to harness soliton compression and fission in integrated photonic platforms. However, implementation in CMOS-compatible platforms has been hindered by the strong nonlinear loss that dominates the propagation of high-intensity pulses in silicon and the low-optical nonlinearity of traditional silicon nitride. Here, we present CMOS-compatible, on-chip Bragg solitons, with the largest soliton-effect pulse compression to date with a factor of ×5.7, along with the first time-resolved measurements of soliton fission on a CMOS-compatible platform. These observations were enabled by the combination of unique cladding-modulated Bragg grating design, the high nonlinearity and negligible nonlinear loss of compositionally engineered ultra-silicon-rich nitride (USRN: $Si_7N_3$).**

# 1. Introduction

Temporal optical solitons, first demonstrated by Mollenauer et al. [1] in optical fiber, were explored for their use in fiber communications to transmit information as they preserve their shape during propagation.[2-4] This dispersionless propagation of the fundamental soliton is due to the balance between self-phase modulation (SPM) and group velocity dispersion (GVD) in the fiber media. Since then, harnessing soliton dynamics has enabled numerous applications and underpins many contemporary nonlinear optical systems, such as pulse compression,[5-9] pulse train generation,[10-12] supercontinuum (SC) generation[13-16] and stable frequency comb generation.[17,18]

Whereas fundamental solitons preserve their shape during propagation, high-order solitons evolve periodically due to the interplay of linear dispersion and Kerr effects in the material. This periodic evolution involves temporal compression, pulse splitting and recovery of the initial soliton pulse shape. A strong perturbation in the system can break this periodicity and initiate soliton fission,[19,20] which manifests as pulse break up and pulse train generation. Under certain conditions this can also be described in terms of modulational instability.[11] Soliton fission, which is also one of the underlying physical effects involved in SC generation,[13] introduces pulse modulations in time, compression and splitting, giving rise to spectral broadening and new peaks in the spectrum.

In 1987, Chen and Mills theoretically predicted a new class of optical soliton associated with the Kerr nonlinearity of the material and the strong dispersion in periodic structures (e.g. a Bragg grating).[21] These so-called gap solitons [21-24] propagate at frequencies within the stop-band induced by the periodic structure, where the transmission is low, and exploit the intensity dependent refractive index to dynamically switch the stop-band from reflection to transmission. The more general class of optical soliton in Bragg gratings is referred to as Bragg solitons,[6,25] in which case the propagation frequencies can lie outside of the stop-band, close to the blue edge where the dispersion is anomalous and the grating transmission can be high, particularly if it is apodized. Fiber-based Bragg soliton propagation has been studied extensively and is the basis of a range of soliton dynamics, such as pulse compression via higher-order soliton evolution and pulse-train generation via modulational instability or soliton fission.[6,11,26] The strong band-edge dispersion allows for soliton dynamics to be studied and harnessed on significantly shorter length scales; experiments to date have been reported in centimeter scale holographically written optical fiber Bragg gratings.[11,23]

However, implementation of Bragg solitons on a chip has been hindered by design, fabrication and material challenges. The periodicity that induces the strong band-edge dispersion brings about practical difficulties in on-chip fabrication processes. Fabrication inaccuracies in periodic structures give rise to phase discontinuities that can compromise the dispersive properties of the grating band edge, thus precluding their usefulness.[27] At the same time, CMOS platforms are constrained by the inherent material properties, which has impeded progress: Silicon waveguides possess non-negligible nonlinear losses and traditional silicon nitride has low optical nonlinearity.

In this paper, we report the observation of Bragg solitons and associated dynamics in a CMOS-compatible integrated device, observing Bragg soliton-induced compression up to a factor of ×5.7. This record soliton-effect compression factor (on a CMOS-compatible platform) and observation of soliton fission is enabled by the large Kerr nonlinearity and absence of nonlinear losses intrinsic to the ultra-silicon-rich nitride platform, and the large GVD induced by the periodicity in the integrated cladding-modulated apodized Bragg grating. We present time-resolved measurements of soliton fission for the first time on a CMOS-compatible platform using frequency-resolved electrical gating (FREG), which allows us to acquire the exact temporal shape and phase information of the output ultrashort pulses and sheds light on this important physical phenomenon. We confirmed our results and elucidated the underlying physics with numerical modeling via the generalized nonlinear Schrödinger equation (GNLSE).

## 2. Material platform and the cladding-modulated Bragg gratings (CMBGs)

We created the ultra-silicon-rich nitride [16,28] platform with a material composition of $Si_7N_3$ to overcome the nonlinear loss limitations found in silicon at the prolific 1.55 μm wavelength region [29,30] while retaining a large nonlinearity. USRN is a CMOS-compatible platform that unlocks an order of magnitude larger Kerr nonlinearity ($n_2$=2.8×10$^{-13}$ cm$^2$W$^{-1}$) compared to stoichiometric silicon nitride in the absence of nonlinear losses. This is made possible via engineering the material's silicon and nitrogen composition to achieve a sufficiently large bandgap to eliminate two-photon absorption. SC generation [16] and high-gain optical parametric amplification [28] were previously demonstrated on this platform. Fabrication details can be found in Supporting Material.

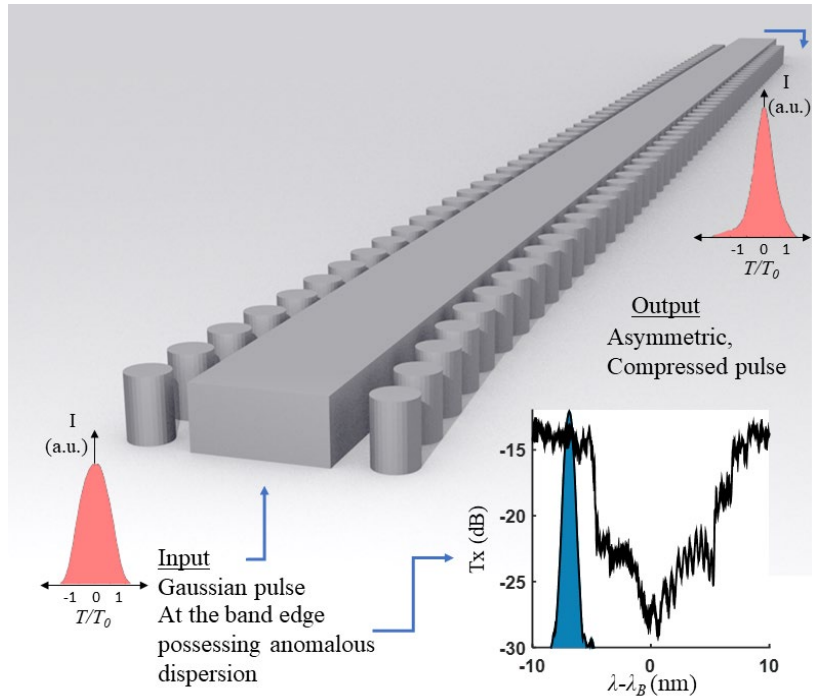

**Figure 1.** 3D schematic of cladding-modulated Bragg gratings (CMBGs). Input and output pulses show the effect of nonlinear pulse propagation along a Bragg grating at wavelengths close to the stop-band for a picosecond pulse.

Experimental time-resolved characterization of Bragg soliton dynamics is performed on cladding-modulated Bragg gratings (CMBGs).[31,32] The 3D schematic of our device which has silica under and over-cladding is given in **Figure 1**. These structures allow dispersion engineering by providing wide flexibility for achieving complex gratings via their tunable parameters such as pillar radius ($r$), gap ($G$), and grating pitch ($\Lambda$). Compared to photonic crystal waveguides (PhCWgs) which are largely implemented using membrane structures, CMBGs require a simple lithographic step with no sacrificial etch and are thus more structurally robust. And compared to Bragg gratings that exploit photosensitivity (e.g. photodarkening), these structures are more stable. The unique cladding-modulated grating design allows for effective apodization in the input and output ends of the grating to minimize sidelobes and ensure low insertion loss at frequencies close the stop-band.[32]

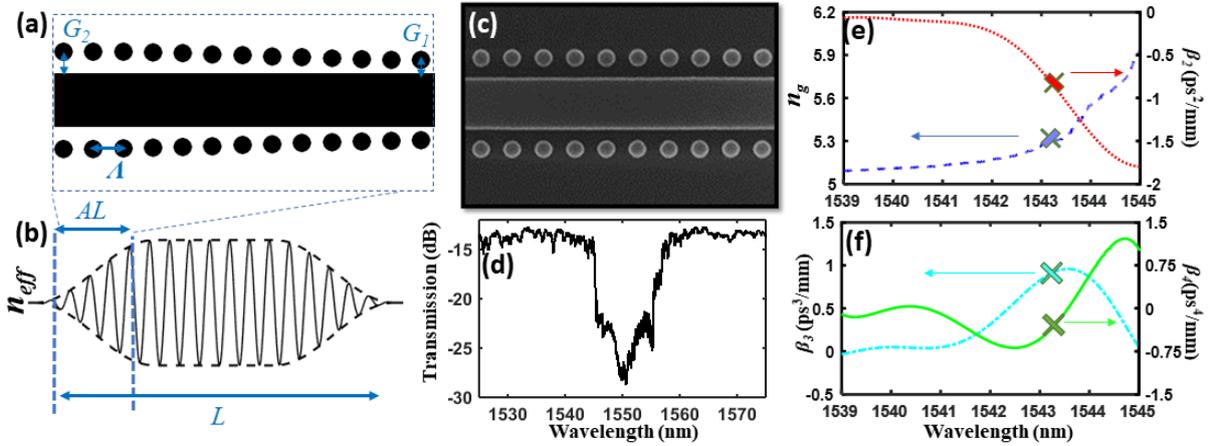

**Figure 2.** Device schematic and optical properties of the CMBG. (a) Top view schematic of the CMBG along its apodization length ($AL$) with its parameters denoted: pitch ($\Lambda$) and gaps ($G_1$ & $G_2$) (b) Effective refractive index ($n_{eff}$) seen by the pulse traveling in the CMBG of length $L$ due to the apodization profile. (c) Scanning electron micrograph (SEM) taken before cladding deposition. (d) Transmission of CMBG showing the band gap of 10 nm. (e) Group index ($n_g$) and group velocity dispersion ($\beta_2$). (f) Third-order ($\beta_3$) and fourth-order dispersion ($\beta_4$) curves.

Apodization can be easily understood through the schematic along the apodization length (AL) shown in **Figure 2a**. The apodization profile is implemented by continuously decreasing the pillar-waveguide gap from $G_2 = 150$ nm to $G_1 = 50$ nm over length $AL=600$ μm according to a raised cosine function:

$$G(z) = (G_2 - G_1) \times \cos^2(\pi x) + G_1 \quad \text{where} \quad x = \begin{cases} \dfrac{z}{2AL} & \text{if} \quad z \leq AL \\ 0.5 & \text{if} \quad AL < z < L - AL \\ \dfrac{z-L}{2AL} & \text{if} \quad z \geq L - AL \end{cases} \quad (1)$$

This modulation of the distance between the pillars and the waveguide changes the effective refractive index ($n_{eff}$) perturbation. Avoiding an abrupt change in $n_{eff}$ eliminates the out of band ripples as the apodization suppresses the grating sidelobes without influencing the Bragg wavelength ($\lambda_B = 2n_{eff}\Lambda$). **Figure 2b** is a schematic of the effective refractive index change seen by the propagating pulse along the grating length $L$ due to this apodization scheme and the placement of pillars. We fabricated a CMBG of total grating length $L = 6$ mm with pitch, $\Lambda = 339$ nm. Considering the apodization length of 600 μm, the grating section with a constant pillar-waveguide gap possesses a length, $L_g = L - 2AL = 4.8$ mm. A scanning electron micrograph of the fabricated device before the deposition of the SiO$_2$ cladding is shown in **Figure 2c**.

The linear transmission characteristics of our grating was first measured using an amplified spontaneous emission source and an inline polarizer to ensure transverse-electric polarization. An optical spectrum analyzer was used to measure the grating's transmission spectrum. **Figure 2d** shows the transmission spectrum of our CMBG with its band gap centered close to 1.55 μm. We measured the group index $n_g$ with a component analyzer that uses an interferometric technique. The dispersion parameters are derived from the polynomial fitting of the group index curve shown in **Figure 2e**. The group velocity dispersion (GVD, $\beta_2$), third-order dispersion (TOD, $\beta_3$) and fourth-order dispersion (FOD, $\beta_4$) curves are shown in Figure 2e and **Figure 2f**. The grating dispersion parameters are extracted to be, $\beta_2 = -0.81$ ps$^2$/mm, $\beta_3 = 0.83$ ps$^3$/mm, $\beta_4 = -0.33$ ps$^4$/mm. Dispersion lengths are given by $L_{Dx} = \dfrac{T_0^x}{\beta_x}$, where $x$ denotes the order of dispersion. Using the experimentally obtained values of $\beta_n$, the corresponding dispersion lengths are calculated to be $L_{D2} = 10.8$ mm, $L_{D3} = 31.3$ mm and $L_{D4} = 23.3$ cm at $\lambda = 1543.1$ nm, considering a pulse full width half-maximum of 4.93 ps ($T_{FWHM} \approx 1.665 \times T_0$). $L_{D2}$

and $L_{D3}$ are sufficiently small that GVD and TOD will have a significant effect on the nonlinear pulse propagation dynamics whereas FOD may be assumed to be of minor importance. Details on CMBG parameters can be found in Supporting Material.

## 3. High-order soliton dynamics in USRN gratings

Soliton propagation in USRN can be modelled by solving the GNLSE which is a good approximation in this regime [33,34] via the split-step Fourier method:

$$\frac{\partial A}{\partial z} = -\frac{\alpha}{2} A + i \sum_{k=2}^{6} \frac{i^k}{k!} \beta_k \frac{\partial^k A}{\partial T^k} + i\gamma_{eff}(|A|^2 A) \qquad (2)$$

The GNLSE assumes that the pulse envelope denoted by $A(z,t)$ is slowly varying. Contributions of chromatic dispersion to the pulse propagation dynamics is included through the second term at the right-hand side. The last term of the **Equation 2** takes into account effects from self-phase modulation (SPM). $\gamma_{eff}$ denotes the effective nonlinear parameter and is calculated as $\gamma_{eff} = \frac{\omega_0 n_2}{\lambda A_{eff}} \left(\frac{n_g}{n_o}\right)^2$, where $\omega_0$ is the frequency of the pulse, $n_g$ is the group index and $n_o$ is the refractive index of the material. The effective area ($A_{eff}$) is engineered to be small in order to maximize the magnitude of $\gamma_{eff}$. Nonlinear losses have been shown to be negligible at optical intensities up to 50 GW/cm$^2$ for the USRN platform,[16] significantly larger than the power levels used in this experiment. Therefore, we only take the linear propagation losses, $\alpha$, into account.

Higher-order solitons are bound states of fundamental solitons that undergo recurrent periodic propagation. The soliton order $N$ is defined as, $N^2 = \frac{L_D}{L_{NL}} = \frac{T_0^2 \gamma_{eff} P_0}{|\beta_2|}$ where $P_0$ refers to the peak power. In the absence of perturbative effects (i.e. TOD, FC and TPA), an $N=1$ soliton propagating through the waveguide preserves its shape over the propagation length; this corresponds to a fundamental soliton. Higher-order solitons exhibit more complex temporal dynamics: they periodically compress, split into multiple pulses, and recover their original shape. When $N$ is an integer number, this periodic evolution occurs at exactly one soliton period, $z_0=\pi/2 \times L_{D2}$.[3] When TOD effects are sufficiently large, the delicate balance of GVD and SPM is disrupted, and this symmetric and periodic soliton evolution ceases to exist. In the absence of TOD, the fundamental solitons comprising a higher-order soliton are bound and behave as a breather, as all of them propagate at the same speed. The presence of TOD breaks that degeneracy and causes the separation of higher-order soliton into its constituents. These separated solitons then travel further apart from each other in the guiding media rather than evolving back into their initial pulse shape at $z=z_0$.

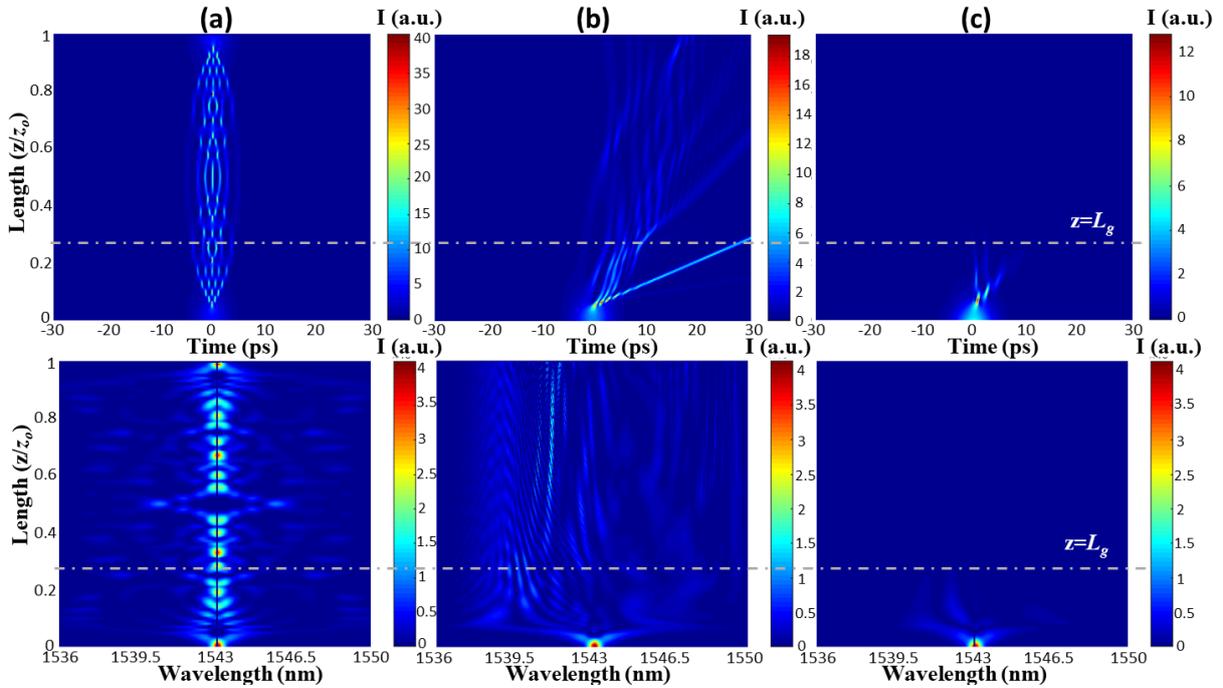

**Figure 3.** Modelled temporal and spectral intensity plots showing the pulse evolution along one soliton period for the soliton order $N=8$. Uniform grating length $L_g$ is marked on the plots. (a) Pulse evolution in the absence of TOD and linear loss. (b) Soliton fission after initial pulse compression when TOD effects are included in modelling. (c) Including all the effects given in Equation 2.

To reveal the underlying physics in the propagation of high-order solitons and their fission, we examine the "ideal" pulse evolution in the gratings where $N=8$ and only SPM and GVD effects are present. As expected, the interplay of the dispersion and nonlinearity results in a complex periodic evolution of the breathers as seen for the $N=8$ case over one soliton period in **Figure 3a**. For the dispersion parameters of our grating, soliton period is $z_0 = 17$ mm. From Figure 3a, we observe that in the absence of TOD, the pulse recovers its original shape at $z = z_0$. Turning on the effects of TOD in our simulations breaks the spectral and temporal periodicity of soliton evolution as seen in **Figure 3b**, as the constituent fundamental solitons move at different speeds due to TOD. This results in the continuous separation of the breathers rather than the oscillations observed in Figure 3a and serves as evidence that the onset of soliton fission is brought about by TOD in our platform. The soliton is not able to restore its initial spectral shape but instead, the separation in time manifests itself in the spectral domain as an overall broadening and formation of several spectral peaks, as well as a blueshift. Finally, we include loss and fourth-order dispersion (FOD) and numerically solve Equation 2. The simulations in **Figure 3c** reveal that the initial input pulse evolves into separate pulse peaks. We may further conclude from the numerical simulations that gratings of uniform length $L_g \geq 4.8$ mm are sufficiently long for soliton dynamics and fission associated with a soliton order of $N = 8$ to be observed.

**4. Time-domain characterization of soliton compression and fission**

For time-resolved characterization, we used a FREG apparatus, shown in **Figure S1**. FREG allows us to retrieve the exact temporal shape and the phase of the ultra-short pulses. [35] The FREG setup measures a set of time-gated spectra using the delayed input pulse and the output of CMBG. This provides spectrograms which we use to extract the pulse intensity and phase via a numerical algorithm. [36] Further details on FREG apparatus can be found in Supporting Material.

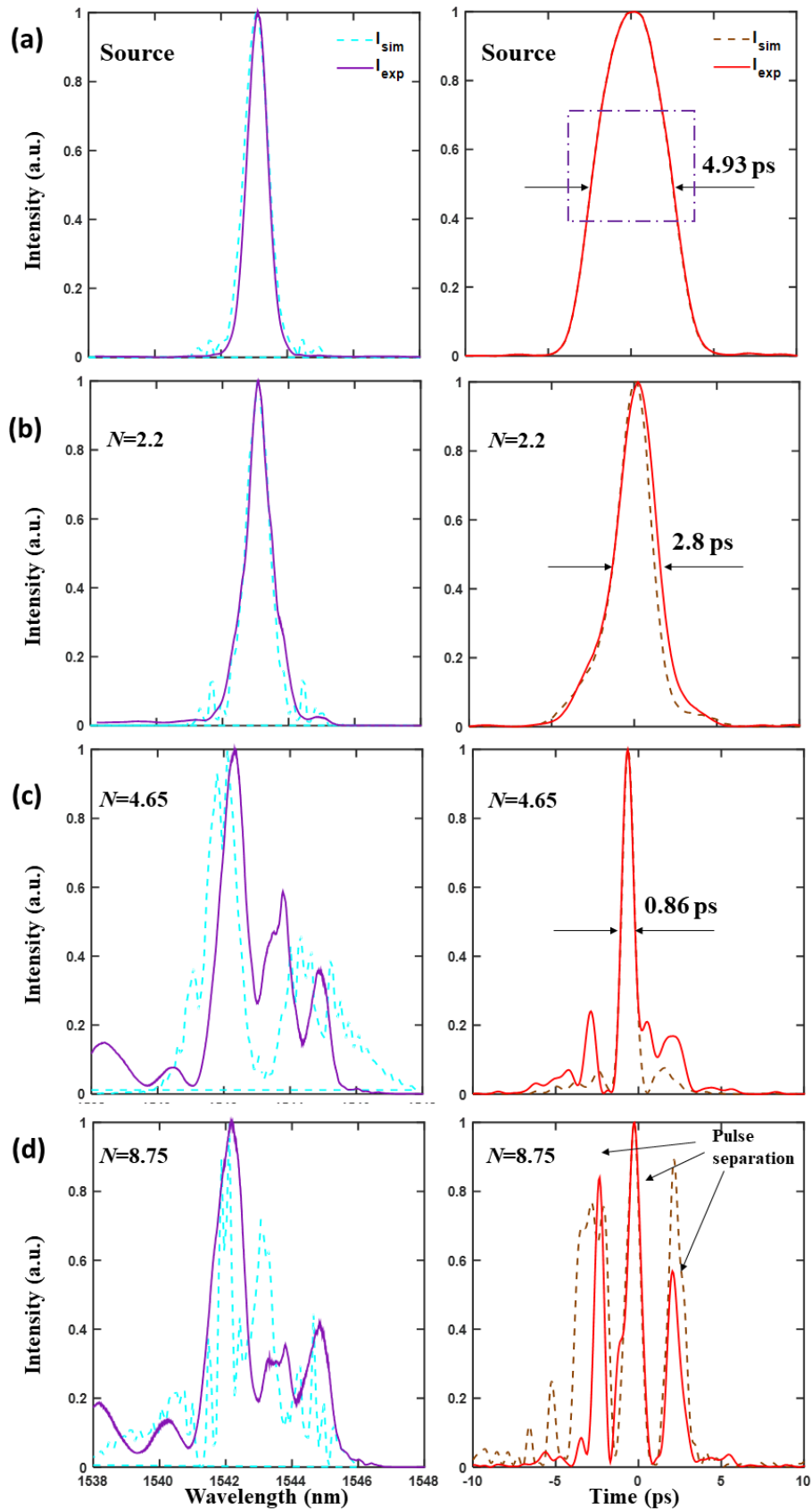

**Figure 4.** Temporal and spectral characterization of the pulses exiting the CMBG. Measured spectrum and temporal profile of (a) the source pulse, soliton evolution for (b) $N$ = 2.2, (c) $N$ = 4.65 and (d) $N$ = 8.75 for $L_g$ = 4.8 mm.

For the temporal and spectral observations of high-order soliton dynamics in CMBGs, picosecond Gaussian pulses are coupled into the gratings using tapered fibers. **Figure 4a** shows the intensity profile of the input pulses used in the experiments, generated by our source setup of cascaded mode-locked laser, wave shaper and Erbium-doped fiber amplifier (EDFA). We input these near-transform limited 4.93 ps pulses with approximately flat phase (as shown in **Figure 5a**) and observe the spectral and temporal evolution of the pulses as the power coupled into the gratings and hence soliton order $N$, increases.

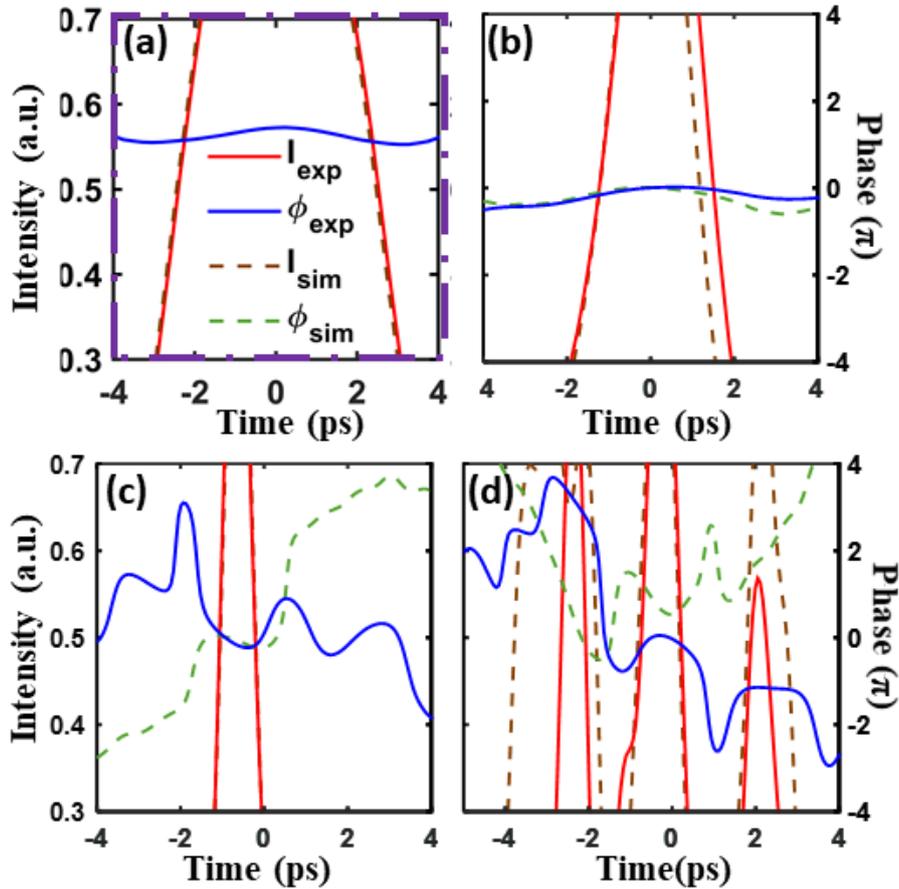

**Figure 5.** The measured phase at the pulse centers (as marked with purple dashed square in Fig. 4. (a)), from right to left; (a) the source and high-order soliton evolution for (b) $N = 2.2$, (c) $N = 4.65$ and (d) $N = 8.75$ for $L_g$=4.8 cm.

For $N = 2.2$ ($P_0$=0.36 W), the pulse compresses with a factor of ×1.8 as seen in **Figure 4b** with a slight broadening in the spectral domain. As the input power is further increased, the pulse gradually compresses up to a large soliton compression factor of ×5.7, exceeding the previous demonstrations of on-chip soliton compression [8,9] for $N = 4.65$ ($P_0 = 1.62$ W) as shown in **Figure 4c**. Spectral broadening further increases and starts showing multiple peaks in both measurements and simulations. When we zoom into the phase at the center of the pulse in **Figure 5c**, we see that both are flat with an asymmetry induced by the grating's TOD.

We increased $N$ by further increasing the coupled power into the grating and observed fission for the soliton order $N=8.75$. Our simulations as well as the measurements show three clear peaks at an intensity exceeding a quarter of the peak intensity. Spectral characterization of the $N=8.75$ ($P_0=5.73$ W) case reveals new spectral peaks on the blue side of the spectrum. It is expected from the nature of Fourier transformation that modulation in the spectral domain will be accompanied by temporal separation. The distinct peaks in the time domain show themselves as the new peaks forming in the frequency domain; a signature of the TOD-induced fission is the new peak formation in the anti-Stokes side. [33] This is clearly seen in the spectral domain in **Figure 4d** at $\lambda=1540$ nm for both modelling and experimental results. As a result of the pulse separation, we see a phase difference of $\pi$ between three consecutive pulses in our phase-resolved measurements, as seen in **Figure 5d**.

As the extent of spectral broadening is increased, the dispersion experienced by newly generated spectral fringes differs from that of the pulse's central wavelength. This results in the variations of widths and amplitudes of fundamental solitons in order to conserve their soliton order. Solitons that have undergone fission do not propagate as a single entity as the constituent fundamental solitons do not propagate with the same velocity. Considering the inherent uncertainties of high-power measurements and very high order nonlinear effects, our modelling and time-resolved measurements, while not in perfect agreement, agree sufficiently well such that the underlying physics of soliton dynamics are well captured. The moderate quantitative deviations can be explained by several factors: rapid change of dispersion at the band edge, divergences from the ideal input pulse shape, polarization tuning and power fluctuations as well as the complexity of modelling nonlinear effects induced by high input peak powers.

We observe through time-resolved measurements that the periodicity of soliton evolution breaks as the constituent fundamental solitons move at different speeds. This process, referred to as soliton fission, occurs here due to TOD perturbation. [19,33] Our modelling shown in Figure 3 - 5 verifies the fission in our cladding-modulated Bragg grating. In addition, analytic formalism from the literature states a threshold $\beta_0$, for TOD to break the symmetry of the propagating pulses and initiate the fission of high-order solitons. $\beta_0$ rapidly decreases with increasing $N$ and the parameter $\beta$ calculated as $\beta = \beta_3/6\beta_2 T_0$ should exceed the threshold $\beta_0$ [19]. $\beta$ provides a means to control soliton interactions with TOD to improve bandwidth in fiber communications. [37] For conditions leading to $N = 3$, we are above the threshold value of $\beta_0$ required to initiate solution fission as $\beta = 0.058$ for our CMBG. [19] We therefore infer that we are also above this threshold for our experimental demonstration of soliton fission where $N = 8.75$ and conclude that the pulse splitting observed for $N = 8.75$ is a result of soliton fission occurring. Furthermore, the fission distance, which is defined by $L_{fiss} \approx L_{D2}/N$, [13] corresponds to $L_{fiss} = 1.23$ mm for the high power measurements ($N = 8.75$) which is much shorter than of our uniform grating length $L_g = 4.8$ mm; for higher powers the minimum grating length to observe fission would be even shorter.

For comparison, we carried out time-resolved characterization for a grating with shorter length, $L=3$ mm, $AL=600$ um and therefore $L_g=1.8$ mm. We observed less asymmetry in the gradual pulse compression up to a factor of ×2.37, but we were not able to observe fission for this grating. **Figure 6a** shows the pulse width vs input peak power for CMBG of total length 3 mm, where the input pulse with full-width half maximum of 6 ps shown in the inset. **Figure**

**6b** shows the results of pulse compression for increasing values of $N$ from this set of experiments. We aimed to operate in the regime where the grating length was shorter than the theoretically predicted length required for fission to be observed. The increased pulse duration compared to our results presented in Fig. 5, results in an increase in $L_{D2}$, from 10.8 mm to 16 mm. This results in an associated increase in the fission length, $L_{fiss}$~1.5 mm for $N$=10.8. Consequently, we believe that the uniform grating length, $L_g$=1.8 mm, is insufficient to observe separation of the breathers in this grating. Conversely, the experimental characterization for this shorter grating reveals that the pulses may be effectively compressed, resulting in optical pulses with better symmetry and lower amplitude soliton pedestals.

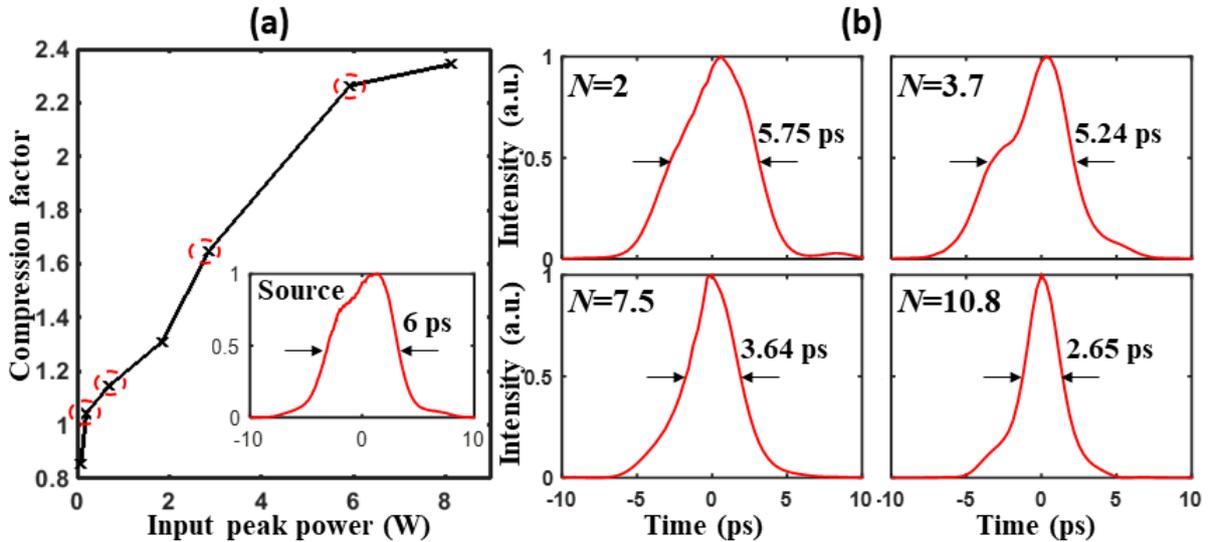

**Figure 6.** Temporal compression using CMBG of total length 3 mm and $L_g$=1.8 mm. (a) Compression factor vs input peak power where the inset shows the Gaussian input pulse of width of 6 ps. (b) Evolution of output compressed pulses for the data points marked on Fig. 6 (a).

## 5. Conclusion

We highlight the importance of our record high-order soliton compression result. Achieving a compression factor of ×5.7 through high-order soliton compression in CMOS-compatible platform is very promising for applications in optical signal processing and ultrafast integrated lasers; even larger compression factors can be achieved by exploiting higher soliton orders while using a shorter grating and larger powers that were not available using our current setup.

We note that on-chip soliton-based pulse compression has been demonstrated in two-dimensional PhCWgs: in GaInP (×5.4 compression) [8] and silicon (×2.3 compression). [9] However, GaInP is not CMOS-compatible and the latter and the latter demonstration was fundamentally limited by strong nonlinear losses from two-photon and free-carrier effects, thus precluding any observation of strong compression or soliton fission. While solitons in PhCWgs have some similarities with Bragg solitons in that they both occur in very short long scales given the strong dispersion of the underlying periodic medium, solitons formed in nonlinear Bragg gratings have some important advantages. Firstly, integrated Bragg gratings are compatible with conventional waveguides, and unlike PhCWgs, [38] do not possess large insertion losses. Further, they also possess reduced propagation losses, and have better mechanical robustness than membrane-based PhCWgs due to their cladded structure. When combined with the USRN platform which possesses a high nonlinear figure of merit, [28] nonlinear pulse propagation in Bragg gratings can be efficiently harnessed for novel functionalities with potential impact in various fields including ultrafast waveform metrology, imaging and optical communications.

Furthermore, by carefully controlling the grating parameters, for example by introducing chirp, [32] high-repetition-rate soliton-trains can also be generated, as previously proposed for fiber Bragg gratings. [39] Using this method, the characteristics of the generated soliton-train, such as repetition rate and pulse width could also be tailored for advanced on-chip light sources at GHz repetition rates and with footprints significantly smaller than fiber-based systems. Our demonstration of soliton fission could herald a new class of on-chip light sources that leverage fission for femtosecond pulse train generation. Moreover, the fission process will enable enhancing the efficiency of SC generation. [40] This work should inspire researchers to engineer and exploit higher-order dispersion in periodic media and to examine nonlinear pulse propagation, and in particular, to elucidate and harness the complex dynamics associated with pulse train generation and soliton fission.

**Acknowledgements**

E.S. acknowledges scholarship funding Singapore International Graduate Award (SINGA) from A*STAR and thanks the Institute for Photonics and Optical Science (IPOS), the University of Sydney Nano Institute and the School of Physics at the University of Sydney for hosting her to conduct the experiments with A.B.R.  D. T. H. T. acknowledges the support of the MOE ACRF Tier 2 grant, National Research Foundation Competitive Research Grant, SUTD – MIT International Design center, Temasek Laboratories grant and the National Research Foundation, Prime Minister's Office, Singapore, under its Medium Sized Centre Program. B.E. acknowledges the support of the Australian Research Council (ARC) Laureate Fellowship (FL120100029) and the Centre of Excellence program (CUDOS CE110001018).

**Supporting Material**

1. **USRN CMBG fabrication**

Fabrication of the CMBGs is performed on a silicon substrate with 10 μm $SiO_2$ thermal oxide layer. USRN with a film thickness of 300 nm is first deposited using inductively-coupled chemical vapor deposition at a low temperature of 250°C. Thereafter, the gratings are defined using electron-beam lithography, inductively coupled plasma etching, followed by 2 μm $SiO_2$ cladding deposition using atomic layer deposition and plasma-enhanced chemical vapor deposition.

2. **Material and CMBG properties**

Linear and nonlinear refractive index of the USRN is given by the parameters $n_0$=3.1 and $n_2$ =2.8×$10^{-13}$ $cm^2$ $W^{-1}$. CMBG design has a waveguide width of 600 nm and thickness of 300 nm. For the grating used in demonstration of large soliton compression and fission CMBG parameters are $r$=100 nm, $\Lambda$=339 nm, $G_1$=50 nm, $G_2$=150 nm, $L$=6 mm, $AL$=600 μm hence $L_g$=4.8 mm. We characterized the CMBGs' propagation loss to be $\alpha$=13 dB/cm using the cut-back method. The effective area of the CMBG is $A_{eff}$=0.26×$10^{-8}$ $cm^2$. Since our grating has a tapered input waveguide length of only 250 μm we did not include the effect of the access waveguide in our simulations.

3. **Time and phase-resolved characterization**

The experimental setup shown in **Figure S1** consists of mode locked laser (Alnair) cascaded with a pulse shaper (Finisar) and Erbium-doped fiber amplifier (HibriLaser) generating near transform-limited 4.93 ps pulses at 1543.1 nm at a 30MHz repetition rate and the FREG apparatus. The pulses are split into two branches by a fiber-coupler, with most of the energy coupled into the CMBG. In the reference branch, the pulse goes through a variable delay, then it is detected by a fast photodiode and transferred to the electronic domain. The electronic signal generated by the photodiode drives the Mach–Zehnder modulator that is used for time-gating the optical pulse output from the CMBG. Finally, using an optical spectrum analyzer, we measured the spectra as a function of delay to generate optical spectrograms. A numerical algorithm (256×256 grid-retrieval errors G<0.005) is used for the deconvolution of the spectrograms to retrieve the pulse intensity and the phase in both the temporal and spectral domain.[9]

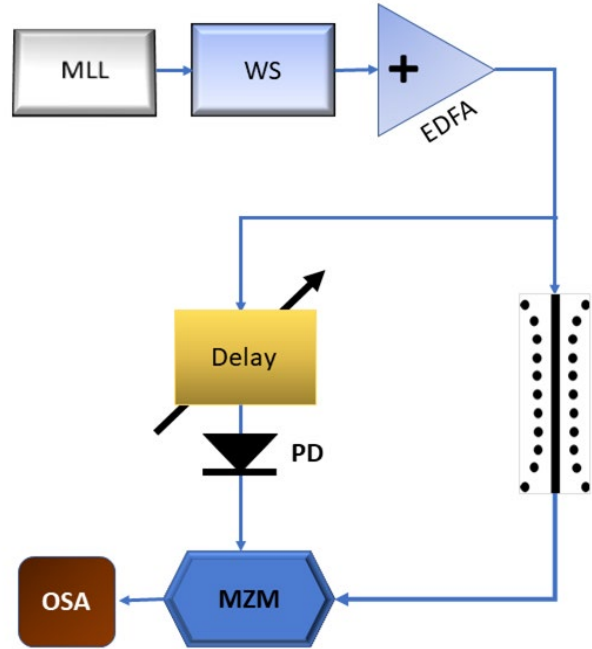

**Figure S1**. Experimental setup consisting of source; mode-locked laser (MLL), wave shaper (WS) and erbium-doped fiber amplifier (EDFA) as well as the frequency-resolved electrical gating apparatus with the CMBG sample, tunable delay arm, ultrafast photodetector (PD), Mach-Zehnder modulator (MZM) and optical spectrum analyzer (OSA).